\documentclass{article}
\usepackage{graphicx}
\usepackage{multicol,caption}
\usepackage[a4paper,
            bindingoffset=0.2in,
            left=1in,
            right=1in,
            top=1in,
            bottom=1in,
            footskip=.25in]{geometry}
\usepackage[backend=bibtex]{biblatex}
\usepackage{amsmath}
\DeclareUnicodeCharacter{2212}{\textendash}
\DeclareUnicodeCharacter{2009}{ }

\usepackage[pdfpagelabels]{hyperref}

\newenvironment{Figure}
  {\par\medskip\noindent\minipage{\linewidth}}
  {\endminipage\par\medskip}

\newcommand{\threepahsecontactline}{3-phase contact line\xspace}

\newcommand{\figref}[1]{Fig.~\ref{#1}}

\newcommand{\etall}{\textit{et~al.}\xspace}

\newcommand{\review}[1]{#1} 
\newcommand{\mytitle}{Confocal Raman microscopy inside sessile multicomponent droplets}
\newcommand{\aperatureangle}{\ensuremath{\gamma}\xspace}
\newcommand{\na}{\ensuremath{N\!A}\xspace}
\newcommand{\focusnew}{\ensuremath{F'}\xspace}
\newcommand{\focusold}{\ensuremath{F}\xspace}
\newcommand{\dropcontour}{\ensuremath{C_{nf}}\xspace}
\newcommand{\refractionangletotal}{\ensuremath{\alpha}\xspace}
\newcommand{\refractionanglelasercone}{\ensuremath{\alpha_{i2}}\xspace}
\newcommand{\incidentanglelasercone}{\ensuremath{\alpha_{i1}}\xspace}
\newcommand{\refractionindexdrop}{\ensuremath{n}\xspace}
\newcommand{\laserconeleft}{\ensuremath{B_l}\xspace}
\newcommand{\laserconeright}{\ensuremath{B_r}\xspace}
\newcommand{\laserconebeams}{\ensuremath{B_i}\xspace}
\newcommand{\laserconeleftrefracted}{\ensuremath{B_l'}\xspace}
\newcommand{\laserconerightrefracted}{\ensuremath{B_r'}\xspace}

\newcommand{\normal}{\ensuremath{N_i}\xspace}
\newcommand{\contactangle}{\ensuremath{\beta}\xspace}
\newcommand{\dx}{\ensuremath{\Delta x}\xspace}
\newcommand{\dy}{\ensuremath{\Delta y}\xspace}
\newcommand{\intersectionleftlasercone}{\ensuremath{I_l}\xspace}
\newcommand{\x}{\ensuremath{x}\xspace}
\newcommand{\y}{\ensuremath{y}\xspace}
\newcommand{\xnorm}{\ensuremath{x^*}\xspace}
\newcommand{\ynorm}{\ensuremath{y^*}\xspace}
\newcommand{\dropheight}{\ensuremath{h}\xspace}
\newcommand{\dropradius}{\ensuremath{r}\xspace}
\newcommand{\xfocusshiftnorm}{\ensuremath{\Delta x^*}\xspace}
\newcommand{\yfocusshiftnorm}{\ensuremath{\Delta y^*}\xspace}

\usepackage{color}
\usepackage[dvipsnames]{xcolor}
\usepackage{siunitx}
\usepackage{xspace}

\title{\textbf{\mytitle}}
\author{Alexander Erb\footnote{Technical University Darmstadt, Institute of Material Science, Physics of surfaces, Peter-Grünberg-Str. 16, 64287 Darmstadt, Germany}, Johanna Steinmann$^*$, Youngeun Lee$^*$, Robert W. Stark$^*$\footnote{Corresponding author email address: robert.stark@tu-darmstadt.de}}

\begin{document}

\maketitle

\section*{Abstract}
Evaporating multicomponent droplets are ubiquitous and appear in a variety of everyday situations and technological applications, including coating, 3D printing, and energy conversion processes. During evaporation, concentration gradients are typically induced, resulting in flows within the droplets. Many of the mechanisms underlying multicomponent droplet evaporation are not fully understood. However, most methods utilize markers that can be surface active and thus affect droplet dynamics. Thus, high-resolution marker-free measurements of concentration gradients in evaporating multicomponent droplets are needed. Raman microscopy can provide such a method. However, as the Raman laser has to undergo a phase transition, it is refracted, leading to a distorted drop contour. In this study, we model refraction in the droplet to better understand the shift in focus leading to geometric distortion. The horizontal and vertical shifts in focus are analyzed, and a modified configuration that enables Raman measurements with a horizontal laser is introduced. For both configurations, the simulated drop shape in the Raman image fits well with the measured shape. While Raman measurements with a vertically incident laser allow the study of the upper part of the droplet, for droplets with large contact angles, horizontal Raman measurements allow for the analysis of the region around the \threepahsecontactline. As an example, a concentration map of an evaporating \qty{4.2}{\micro\liter} glycerol/water droplet is presented. The results of this study can be used as a guide for Raman microscopy measurements of droplets. These findings contribute to understanding droplet dynamics during evaporation and provide a basis for developing novel printing, cleaning, and energy conversion technology applications. \newline\newline
\textbf{Keywords:} Quantitative Raman microscopy, Droplets, Evaporation, Concentration gradients, Refraction
\vspace{1cm}

\begin{multicols}{2}

\section{Introduction}\label{sec1}
The evaporation of sessile multicomponent drops is a ubiquitous phenomenon with numerous practical applications in printing \cite{abdolmaleki_dropletbased_2021, kang_3d_2016, ng_controlling_2021, weygant_drop--demand_2023, lim_self-organization_2008}, coating \cite{eslamian_development_2018}, and cooling \cite{di_marzo_evaporative_1993, chakraborty_analysis_2017}.
\newline
During droplet evaporation, inner droplet flows are induced, which determine the droplet behavior and the pattern of the deposit remaining on the substrate \cite{lohse_physicochemical_2020}. A prominent example is the coffee strain effect, wherein an outward capillary flow near the substrate transports coffee particles toward the \threepahsecontactline (the line where the solid, liquid, and gas phases meet), resulting in the formation of a coffee ring after evaporation  \cite{thayyil_raju_evaporation_2022, popov_evaporative_2005, larson_transport_2014}. The reason for these flows is the nonuniformly distributed evaporation rate across the droplet surface. This causes concentration and surface tension gradients, which induce inner droplet flows \cite{lohse_physicochemical_2020, diddens_competing_2021, diddens_non-monotonic_2024}. 
For sessile droplets with contact angles less than \ang{90}, the highest evaporation rate occurs at the \threepahsecontactline \cite{diddens_competing_2021, diddens_non-monotonic_2024, wang_wetting_2022, hu_evaporation_2002, deegan_capillary_1997, deegan_contact_2000, kim_direct_2018}. The component with higher vapor pressure undergoes preferential evaporation, leading to the accumulation of substances with lower vapor pressures close to the \threepahsecontactline. This process also causes a surface tension gradient. Within the droplet, natural convection causes the denser component to sink \cite{diddens_competing_2021}. The combined effects of these two mechanisms result in the formation of concentration gradients within the droplet, extending in both the vertical and horizontal directions. By measuring the concentration gradients, conclusions about the prevailing physical processes can be drawn \cite{bell_concentration_2022, erb_visualization_2025}. 
%
\newline
In previous studies, the concentration gradients in evaporating droplets were measured using aggregation-induced emission \cite{cai_aggregation-induced_2017}, \review{laser-induced fluorescence \cite{maqua_composition_2007, kim_direct_2018}, infrared spectroscopy \cite{innocenzi_evaporation_2008}}, and refractometry \cite{wilms_composition_2007}. However, most of these methods use markers, and recently, Diddens \etall reported that even small amounts of a surface-active substance (such as markers) can completely change droplet dynamics \cite{diddens_non-monotonic_2024}. Therefore, marker-free high-resolution measurements of concentration gradients in evaporating droplets are necessary. 
\newline
Raman microscopy is a marker-free, well-established method for noninvasive composition measurements of the interior of droplets. In the past, Raman microscopy was used to analyze the composition of levitating droplets \cite{preston_characterization_1985, thurn_structural_1985, fung_raman_1988, esen_raman_2004}, droplets within living cells \cite{nan_vibrational_2003, zhang_quantification_2017, majzner_raman_2014}, droplets in microfluidic cells \cite{cecchini_ultrafast_2011, cristobal_-line_2006}, droplets in aqueous solutions \cite{neofytos_situ_2024}, and in sessile droplets \cite{bell_concentration_2022, avni_single-droplet_2022, erb_visualization_2025, rostami_coalescence_2025}. Furthermore, Raman microscopic investigations of deposits after droplet evaporation have become an established method \cite{guerrini_surface-enhanced_2021, zhang_raman_2003, gao_label-free_2022, kocisova_analytical_2024}. As demonstrated by Goa \etall, the analysis of blood serum deposits can yield valuable insights into the presence of cancerous cells \cite{gao_label-free_2022}.
\newline
Most of these studies have assumed a homogeneous concentration distribution inside the droplet and thus have acquired only one Raman spectrum per droplet. However, as mentioned above, this is not always the case. Some studies have also considered spatial composition.  Neofytos \etall investigated fat droplets in aqueous solutions and resolved the crystallization process spatially \cite{neofytos_situ_2024}. Majzner \etall locally resolved the composition of lipid droplets in liver cells \cite{majzner_raman_2014}. Bell \etall investigated the concentration distribution within evaporating sessile solvent droplets and mentioned the shifted focus due to the refraction of the Raman laser at the droplet surface \cite{bell_concentration_2022}. 
\newline
In this study, we introduce a confocal Raman microscopy approach for studying the interior of sessile droplets that accounts for refraction at the liquid vapor interface and discuss the reliability of the measurements. 
Furthermore, we discuss horizontal and vertical scanning. For horizontal scanning, a \ang{45} mirror positioned within the beam path allowed the measurement with a horizontal laser. This can reduce refraction effects and enable Raman imaging  close to the \threepahsecontactline. This method was employed to analyze the concentration distribution at the \threepahsecontactline in an evaporating \qty{4.2}{\micro\liter} \qty{10}{mol\percent} glycerol/water droplet. As anticipated \cite{diddens_competing_2021}, preferential water evaporation at the \threepahsecontactline was found, causing an accumulation of glycerol. 

\section{Materials and Methods}
\subsection*{Raman Microscopy}
A confocal Raman microscope (alpha 300R; Witec GmbH, Ulm, Germany) with a green laser (\qty{532}{\nano\meter}, Nd:YAG) operated at \qty{17.5}{\milli\watt} laser power was used for all the measurements. Nikon 10×/0.25 and Nikon E Plan 20x/0.40 objectives were used.  Each spectrum was postprocessed by cosmic ray removal and background subtraction using a rolling ball algorithm. The integral was calculated as the Riemann sum of a given range. For all the measurements, a perfluorodecyltrichlorosilane-coated glass substrate was used. 
\subsection*{Refraction modeling}
The refraction at the liquid‒vapor interface was modeled using a custom Python script. The geometrical operations were performed with the Shapley package \cite{Gillies_Shapely_2025}. The script is available at TU-
datalib \cite{tudatalib}. All simulations presented in this study are based on a water droplet with a refractive index $\refractionindexdrop=1.33$ and an objective with a numerical aperture $\na=0.25$. 
\subsection*{Chemicals}
Glycerol (CAS 56-81-5, \qty{>99.5}{\percent}) was purchased from Sigma Aldrich and used as received. Purified water was obtained from a Milli-Q purification system (Merck KGaA, Darmstadt, Germany).

\section{Results and Discussion}\label{sec3}
During Raman microscopic measurements, where the laser must undergo a phase transition, the laser experiences refraction at the interface. This alters the position of focus so that the intended and actual positions of focus differ \cite{everall_confocal_2000, everall_modeling_2000}. According to Bruneel \etall, when transitioning from an optically thinner medium to a flat optically denser medium, the new focal point is located deeper within the medium \cite{bruneel_depth_2002}.
\begin{Figure}
\includegraphics[width=200pt]{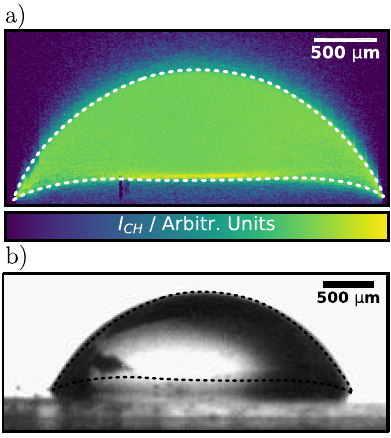}
\captionof{figure}{Raman CH integral image of a sessile \qty{4}{\micro\liter} glycerol droplet and corresponding side image of the same droplet. a) The Raman image was acquired with a vertically incident laser at \qty{17.5}{\milli\watt}, \qty{0.04}{\second} integration time and a resolution of \qtyproduct{10x10}{\micro\meter}. The drop contour is marked with a white dashed line. b) Side image of the same drop. The drop contour shown in the Raman image is marked with a black dashed line. The drop contour of the Raman image does not fully match the drop contour \label{Fig: DropletCrossSection}}
\end{Figure}
%
In \figref{Fig: DropletCrossSection}a, the Raman CH intensity map of a \qty{4.2}{\micro\liter} glycerol droplet cross-section is presented, and the corresponding side view of the drop is shown in \figref{Fig: DropletCrossSection}b. Glycerol is particularly well suited for such studies because evaporation can be neglected because of its low vapor pressure. As shown in \figref{Fig: RamanSpectra}, the integral of the C--H stretching vibrations of glycerol (at approximately \qty[per-mode = power]{2800}{\per\centi\meter}) can be used to characterize the position of the droplet on the glass substrate. The intensity map was recorded using a vertical laser from above. The drop contour in the Raman image is indicated as a white dotted line. The same contour is illustrated in the photograph in \figref{Fig: DropletCrossSection}b as a black dotted line. The outer contours are well fitted. However, in the droplet center, the contours differ. In the Raman image, the center of the glycerol droplet appears to float. The droplet seems to form a dome-like structure with a convex structure in the center.
\begin{Figure}
\centerline{\includegraphics[width=170pt]{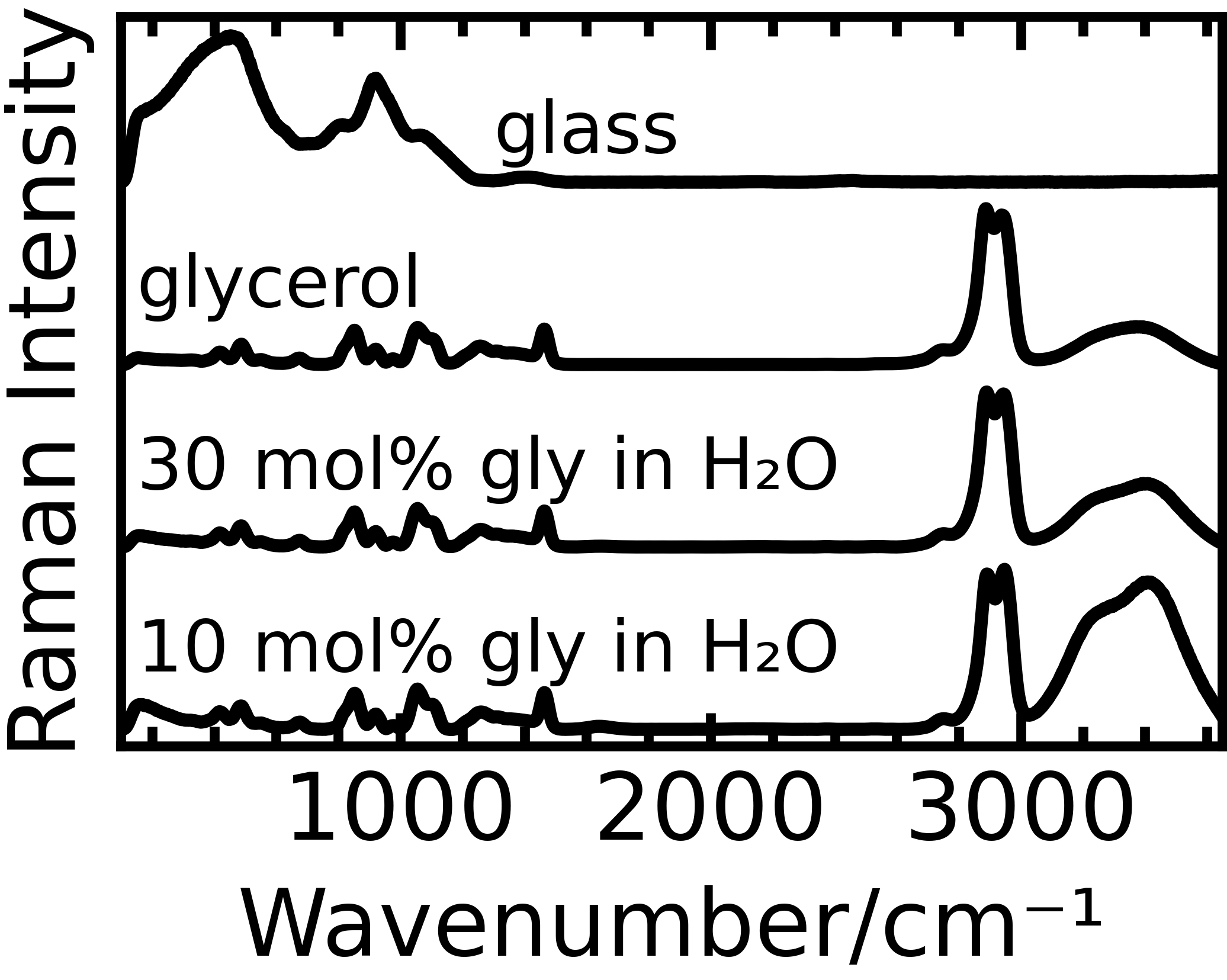}}
\captionof{figure}{Raman spectra of all the materials and fluids used in this study. Spectra were recorded at \qty{17.6}{\milli\watt} with an integration time of \qty{0.5}{\second} on a 10x/0.25 objective. The average of \num{10} spectra smoothed with a moving average filter of size \num{10} is shown.  \label{Fig: RamanSpectra}}
\end{Figure}
%
The reason for this geometric distortion is the refraction of the laser beam at the droplet surface. Because this effect depends on the local curvature of the droplet, the shift in focus is a function of the intended position of focus within the droplet. To gain a more thorough understanding of the shift in focus caused by refraction on the surface of a droplet, simulations are necessary.
\subsection{Focus shift}
To calculate the shift in focus for a sessile droplet caused by refraction, the aperture angle of the objective's opening angle \aperatureangle must first be determined. For an air-suspended droplet measured with an objective with numerical aperture \na, the opening angle of the objective is given as
\begin{align*}
\aperatureangle = \arcsin{(\na)}.
\end{align*}
According to Everall \etall, the new position of focus \focusnew can be calculated as the intersection of the laser cones refracted at the surface \cite{everall_confocal_2000, everall_modeling_2000}. The setup is illustrated in \figref{Fig: RefractionSketch}. In contrast to the flat substrates studied by Everall \etall and other research groups \cite{everall_confocal_2000, everall_modeling_2000, gallardo_confocal_2007, bruneel_depth_2002, baldwin_confocal_2001}, the local curvature of the droplet surface causes symmetry breaking. Thus, refraction of the left and right laser cone beams \laserconeleft and \laserconeright must be considered separately. To calculate the refraction angle \refractionanglelasercone on each side for a given incident angle \incidentanglelasercone in a droplet with refractive index \refractionindexdrop, Snell's law can be used 
\begin{align*} 
\refractionanglelasercone = \arcsin{\big(\sin(\incidentanglelasercone) \cdot \refractionindexdrop^{-1}\big)}.
\end{align*}
\begin{Figure}
\centerline{\includegraphics[width=200pt]{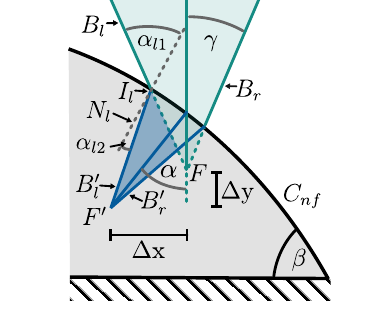}}
\captionof{figure}{Schematic representation of the refraction of the confocal laser cone on a curved droplet surface. The drop with the drop contour \dropcontour and contact angle \contactangle is shaded in gray. The contour can be extracted from side images, which, as it has been shown, do not interfere with the Raman measurement \cite{erb_visualization_2025}. Because of the refraction, the original focal point \focusold is shifted by the distances \dx and \dy to the point \focusnew. The laser cone sides \laserconeleft and \laserconeright (green lines) are refracted toward the center of the drop. The refracted laser cones are labeled \laserconeleftrefracted and \laserconerightrefracted respectively (blue solid lines). For the left laser cone side, the normal \normal is shown as a gray dashed line. The point of intersection with the drop contour \dropcontour is labeled \intersectionleftlasercone. The incident angle of the left laser cone relative to the normal is labeled \incidentanglelasercone, and the corresponding refraction angle is \refractionanglelasercone. The total refraction angle between the original position of focus \focusold and the new position of focus \focusnew is labeled \refractionangletotal.}
\label{Fig: RefractionSketch}
\end{Figure}
To compute the new focal point, to the drop contour must be imported, normalized, and fit. The fitting procedure ensures that the resolution of the drop contour is sufficient for the numerical simulation \cite{xu_algorithm_2014}. The intersection of the drop contour \dropcontour with the two laser cone beams \laserconebeams must first be determined. At this position, the normal \normal is calculated and the incident angle \incidentanglelasercone is determined. Using Snell's law, the refraction angle \refractionanglelasercone can be calculated. The new focal point \focusnew is computed as the intersection of the refracted rays of the laser cone. The algorithm is presented as a pseudocode in \figref{Pseudocode}. To determine the shift in focus for multiple focal points (as needed for a Raman image), the outlined procedure is repeated iteratively. The corresponding Python code is available on TU-datalib \cite{tudatalib}. 


\begin{Figure}
\centerline{\includegraphics[width=200pt]{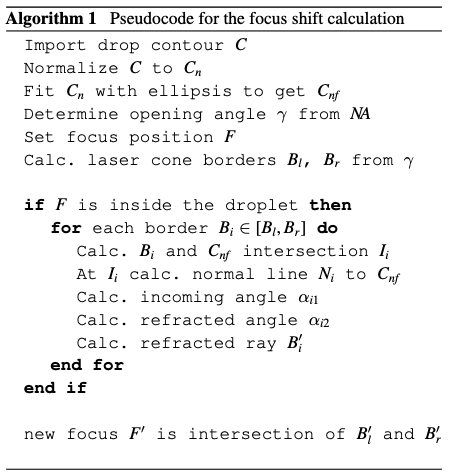}}
\captionof{figure}{Pseudocode for the focus shift calculation}
\label{Pseudocode}
\end{Figure}
\subsection{Vertical Raman scans}
The composition of sessile multicomponent droplets has been investigated several times using a vertically incident laser \cite{bell_concentration_2022, erb_visualization_2025}. However, a detailed discussion of the shift in focus caused by refraction is still needed.
\newline
For the shift-in-focus calculation, the \x and \y positions within the drop were normalized by the drop height \dropheight and radius \dropradius, respectively
\begin{align*} 
\xnorm =& \frac{\x}{\dropradius} \\
\ynorm =& \frac{\y}{\dropheight}.
\end{align*} 
In \figref{Fig: VerticalRefraction}, the shift in focus within the drop is indicated by arrows. An example Raman laser ray is shown as a black arrow. Since the drop (shown in gray) can be assumed to be symmetrical, only the right side is shown. Depth scans into the droplet at various horizontal positions, ranging from the droplet center at $\xnorm = 0$ to close to the \threepahsecontactline at $\xnorm = 0.9$, are presented.
\begin{Figure}
\centerline{\includegraphics[width=200pt]{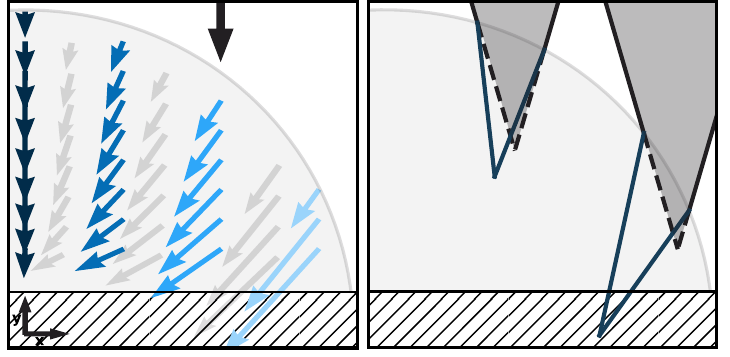}}
\captionof{figure}{Shift in focus at various positions within the drop when a vertical laser is used. The drop is shaded in gray. All arrows start at the intended position of focus and end at the shifted position of focus. Some positions of focus are marked with blue arrows. Their shift in focus is analyzed in more detail in \figref{Fig: VerticalFokusshift}. On the right, the laser cones are shown as two example positions of focus. The refracted laser cone beams are shown in blue. \label{Fig: VerticalRefraction}}
\end{Figure}
For focus positions closer to the \threepahsecontactline, the shift in focus increased. This applies to both the horizontal and vertical shifts. With increasing penetration depth, the shift in focus in the  \x direction increases. However, the shift in focus in the \y direction initially increases until it decreases again at a certain depth. Generally, the greater the difference in slope on the drop surface at the point of intersection with the confocal laser cone is, the greater the shift in focus. This explains the geometric distortion of the Raman image in \figref{Fig: DropletCrossSection}a. The shift in focus due to refraction occurring close to the \threepahsecontactline is so pronounced that only measurements at the droplet surface are possible. This prevents a thorough study of the concentration distribution close to the \threepahsecontactline. In the center of the droplet, no horizontal shift was found. The droplet's center acts as a mirror axis; thus, both laser cones' sides are refracted in the same but mirrored manner. 
\begin{Figure}
\centerline{\includegraphics[width=200pt]{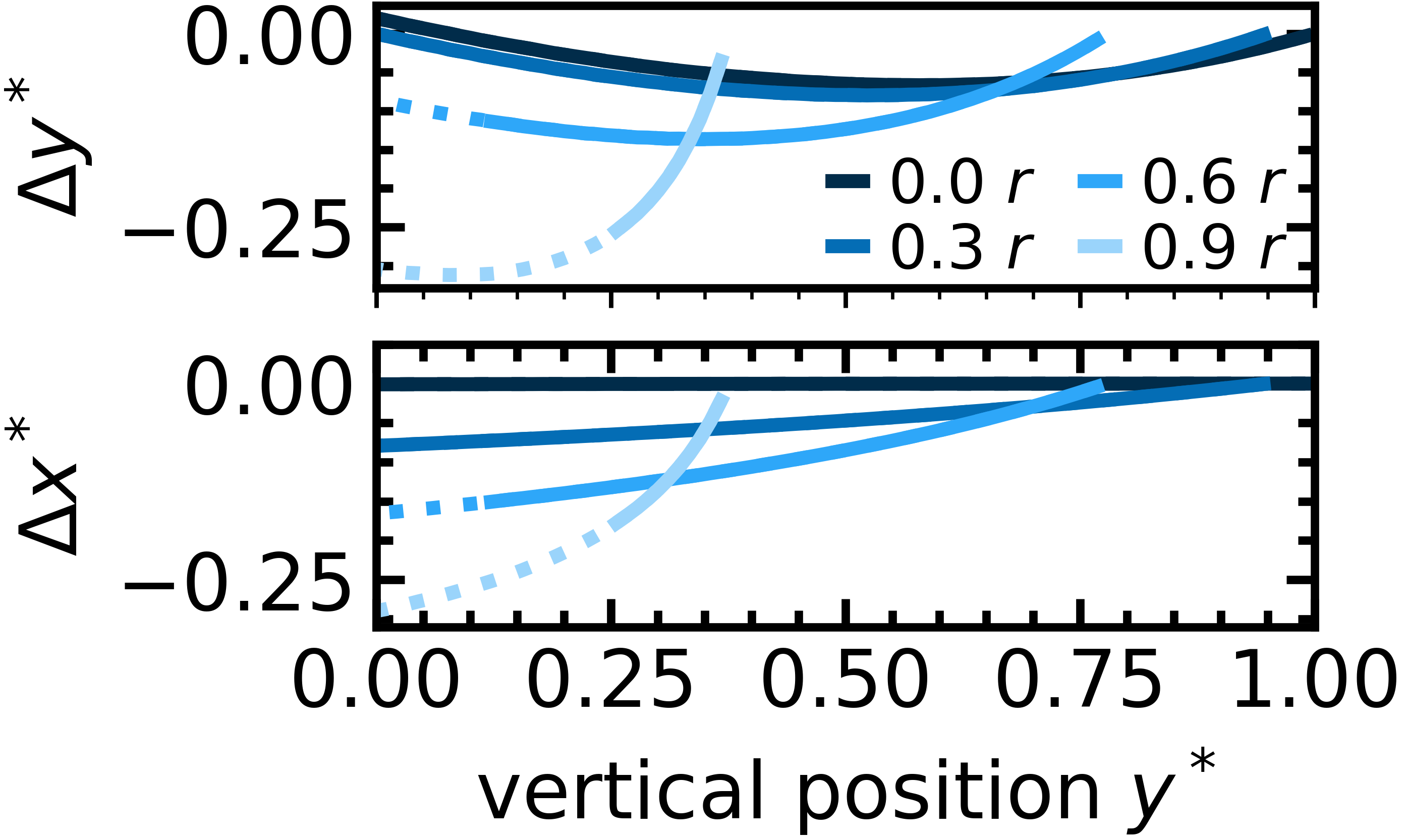}}
\captionof{figure}{Horizontal \xfocusshiftnorm and vertical \yfocusshiftnorm shifts in focus when the drop is scanned vertically as a function of the vertical position \ynorm. The positions closer to the \threepahsecontactline are shown in light blue. Focal positions that are refracted to positions inside the droplet are shown as solid lines. Positions of focus that are refracted into the substrate are shown as dashed lines. \label{Fig: VerticalFokusshift}}
\end{Figure}
%
To examine the shift in focus of a vertical laser in the \x and \y directions more thoroughly,  \xfocusshiftnorm and \yfocusshiftnorm were analyzed for the four positions marked with blue arrows in \figref{Fig: VerticalRefraction}. Positions closer to the \threepahsecontactline are presented as lighter shades of blue. The shift in focus is shown as a function of the intended focus’s vertical position \ynorm. Positions that are refracted into the substrate are indicated by dashed lines. The results are shown in \figref{Fig: VerticalFokusshift}. 
\newline
The shift in focus in the \x-direction \xfocusshiftnorm decreases as the position of focus approaches the center of the droplet, eventually vanishing completely at the droplet's center. When scanning deeper into the droplet, the horizontal shift in focus increases, with the shift direction consistently oriented toward the droplet’s center. Close to the surface, no shift in focus was found.
\newline
The vertical shift in focus also increases with increasing penetration depth. The closer the focal point is to the \threepahsecontactline, the more pronounced the effect. Thus, measurements near the \threepahsecontactline are constrained to the drop surface, as focal points deeper within the droplet shift into the substrate. Interestingly, the shift in focus in the \y-direction \yfocusshiftnorm does not increase continuously with increasing penetration depth. Following an initial increase, the curve reaches its maximum at approximately half the scan depth and subsequently decreases. This phenomenon can be attributed to the higher gradient of the drop contour at the point of intersection with the laser cones.
\newline \newline
To verify our results and investigate the geometric distortion of the droplets shown in \figref{Fig: DropletCrossSection} in more detail, the expected drop shape in the Raman image was calculated. For this purpose, the measurement positions whose shifted focus \focusnew was predicted to still be inside the droplet are shown in blue. To ensure the transferability of the findings to other studies, the procedure was repeated for four drops of different contact angles \contactangle. The results are presented in \figref{Fig: DropShapeVerticalScan}.
\newline
For all contact angles, the lower border of the measurable drop area is S-shaped and convex at the center of the drop and at the \threepahsecontactline.  As the contact angle decreases, the S shape loses its curvature and rotates from the center of the drop toward the \threepahsecontactline. By the end of this rotation, only a parabola remains. Thus, while for high contact angles the center of the drop can be resolved, refraction effects prevent measurements at the \threepahsecontactline. For smaller contact angles, refraction hinders measurements at the center of the droplet. However, owing to the high local gradient close to the \threepahsecontactline, , focal points deeper within the droplet are strongly refracted so that only near-surface measurements are possible.
\begin{Figure}
\centerline{\includegraphics[width=180pt]{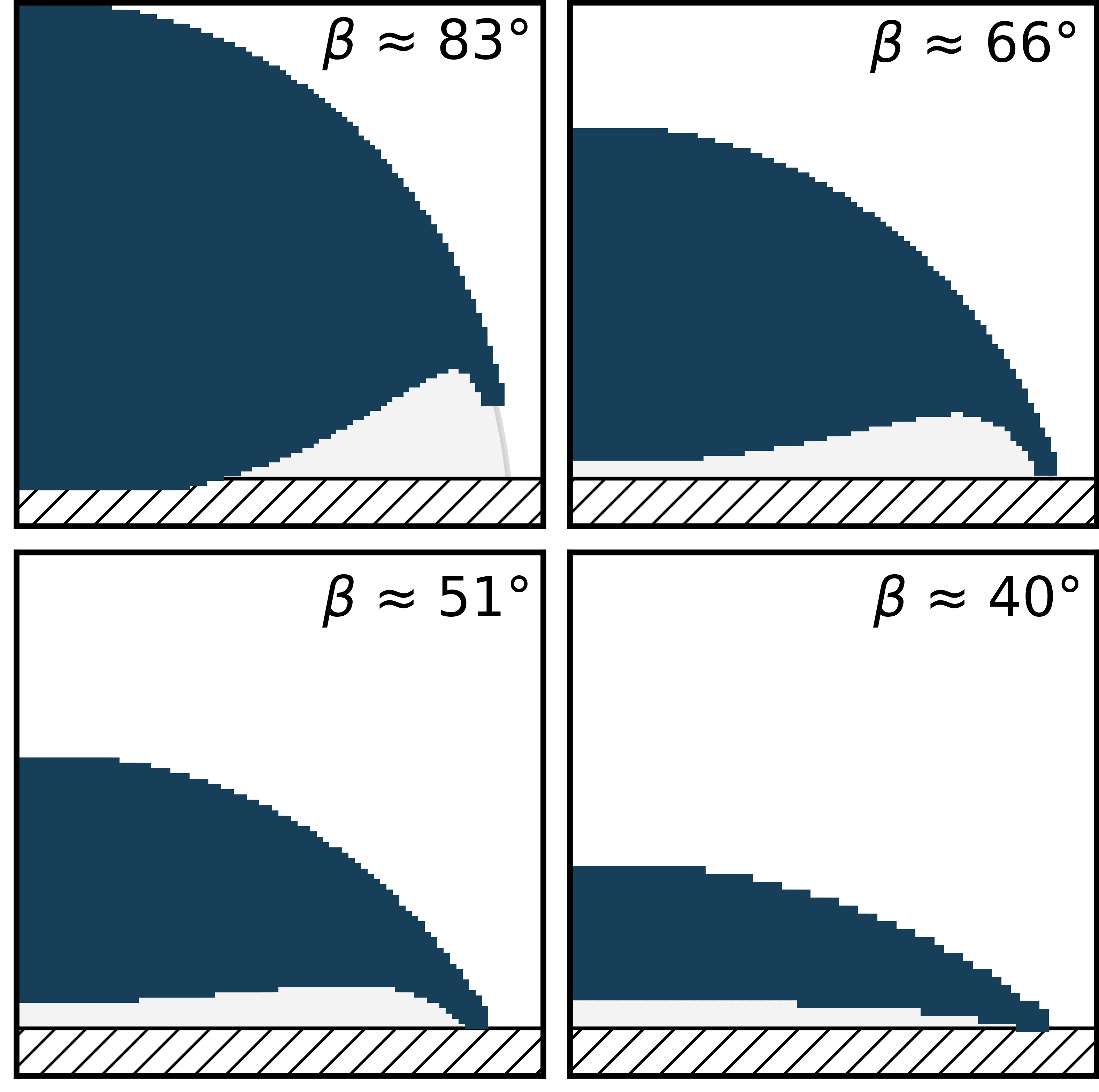}}
\captionof{figure}{Calculated drop shapes in the Raman image for drops with different contact angles, as indicated in the top right corner. The focal points whose focus did not shift into the substrate are marked in blue. The dome shape with a convex center shown in \figref{Fig: DropletCrossSection} is visible in the calculated drop shapes for $\beta=\ang{66}$ and $\beta=\ang{51}$. \label{Fig: DropShapeVerticalScan}}
\end{Figure}
\subsection{Horizontal Raman scans}
Because the preferential evaporation of one of the components is the starting point for numerous processes in the droplet, the area around the \threepahsecontactline is particularly important for understanding the droplet dynamics. As shown in \figref{Fig: DropShapeVerticalScan}, for a vertical laser refraction hinders a detailed study of the concentration distribution at this location. Thus, a \ang{45} mirror was used to redirect the vertical laser beam. In this configuration, a vertical shift in focus can be achieved by shifting the laser position horizontally. The CH-integral Raman image of a \qty{4}{\micro\liter} glycerol droplet near the \threepahsecontactline is shown in \figref{Fig: HorizontalGlyDroplet}. The laser path and the \ang{45} mirror are shown in the image. The droplet is visible as a region of high CH integrals. Starting from the \threepahsecontactline, a region of lower CH integrals increases monotonically toward the center of the droplet. Unlike in the vertical measurement, the lower contour of the droplet blurred in the Raman image.
\begin{Figure}
\centerline{\includegraphics[width=180pt]{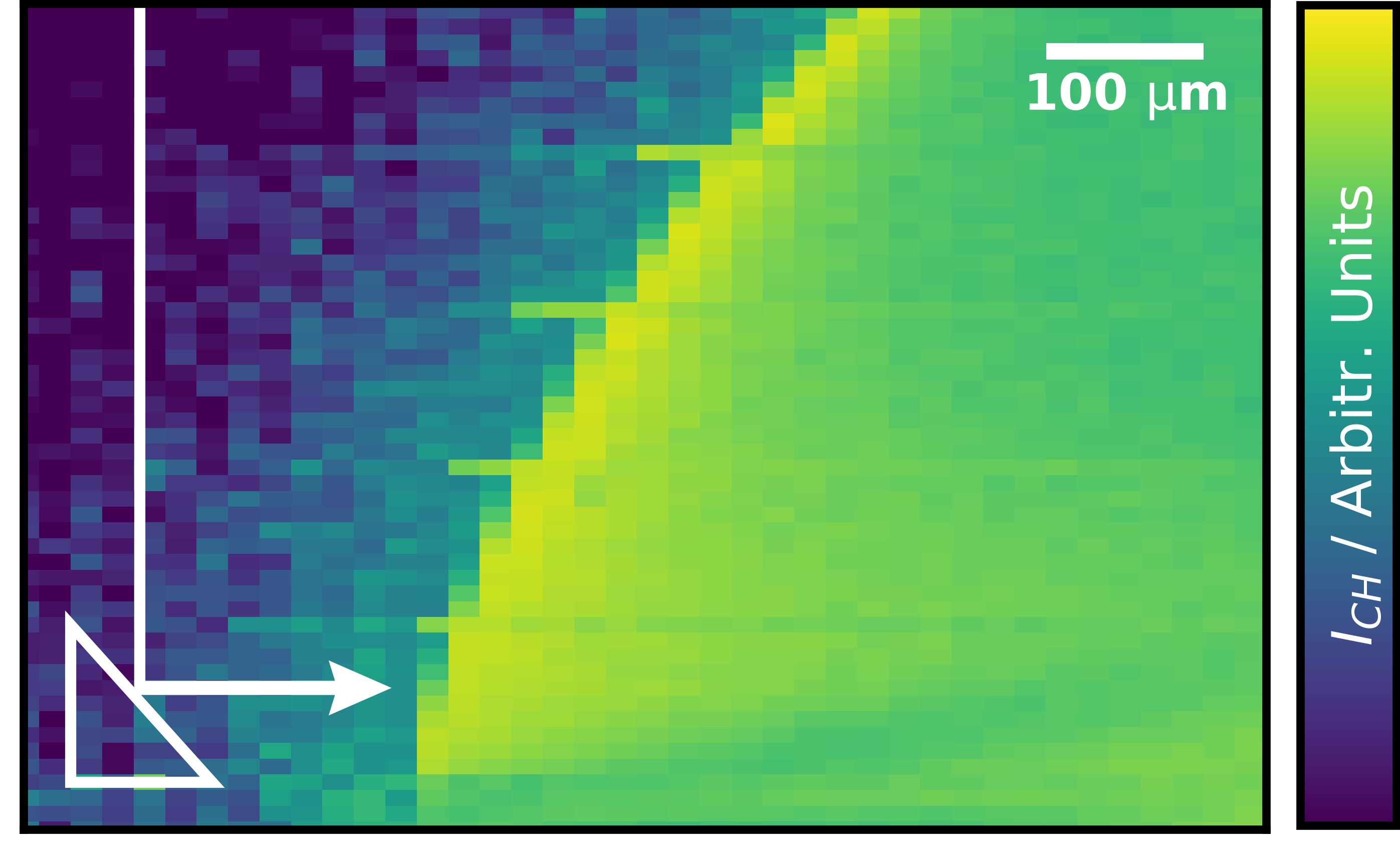}}
\captionof{figure}{C--H Raman image of the lower left side of the droplet measured in the horizontal configuration. A \qty{4}{\micro\liter} glycerol droplet was analyzed using a 10x/0.25 objective at \qty{17.5}{\milli\watt}, with a \qty{0.5}{\second} integration time and a resolution of \qtyproduct{9.7x20}{\micro\meter}. The mirror used to deflect the laser and the laser path are indicated in white in the image. The drop is visible as an area of high integral values (shown in yellow). From the \threepahsecontactline contact line to the center of the droplet, a monotonically increasing band of lower intensity was observed.  \label{Fig: HorizontalGlyDroplet}}
\end{Figure}
To better understand the refraction effects during horizontal Raman measurements, the shift in focus was calculated. It is illustrated as arrows in \figref{Fig: HorizontalRefraction}. Analogous to the vertical measurement, for all positions within the droplet, the focus shifts toward the center of the droplet. The shift increases for higher vertical positions and for increasing penetration depths. However, another effect is visible here: For horizontal scans close to the substrate (light green arrows in \figref{Fig: HorizontalRefraction}), the simulation fails to calculate the refracted focus for positions deep in the droplet. Because the new focus is calculated using the outer beams of the confocal cone and the lower beam is interrupted by the substrate, the simulation cannot determine the new focus. However, in practice, the remaining part of the laser cone continues to penetrate the droplet. Thus, measurements close to the substrate are possible, albeit with less intensity. This explains the strip of lower intensity that approaches the center of the drop in \figref{Fig: HorizontalGlyDroplet}.
\begin{Figure}
\centerline{\includegraphics[width=200pt]{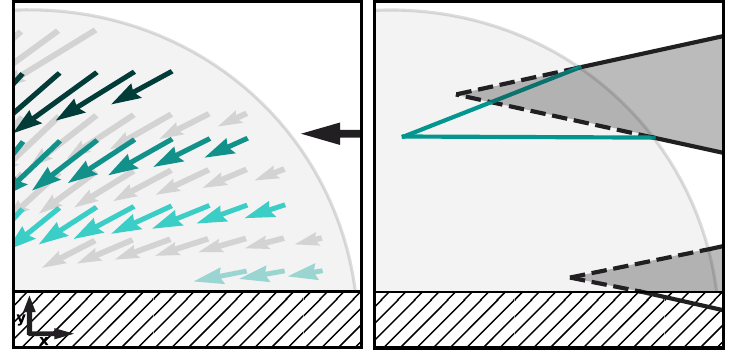}}
\captionof{figure}{Shift in focus within the drop during the measurement with a horizontal laser. The focal points marked in green are analyzed in more detail in \figref{Fig: HorizontalRefractionCurves}. The horizontal laser is indicated schematically as a thick black arrow. The laser cones for two different positions in the droplet are shown on the right. The parts of the laser cones that can reach the intended point of focus are filled in gray. The refracted laser cone beams are shown in green.   \label{Fig: HorizontalRefraction}}
\end{Figure}
%
To investigate the shift in focus during horizontal Raman measurements, the scan positions marked with green arrows in \figref{Fig: HorizontalRefraction} were evaluated. The results are shown in \figref{Fig: HorizontalRefractionCurves}. For higher vertical positions, the horizontal shift in focus increases. The same trend was found for increasing penetration depth. For measurement positions close to the substrate (light green curve), the shift in focus cannot be determined. Thus, the light green curve breaks off at approximately $\xnorm=0.7$, as the laser cone is blocked by the substrate. For the vertical scans, the shift in focus is proportional to the difference in curvature at the entry points of the upper and lower laser cone beams. Since this difference increases with the vertical position, the shift in focus also increases for higher positions in the drop.
\begin{Figure}
\centerline{\includegraphics[width=200pt]{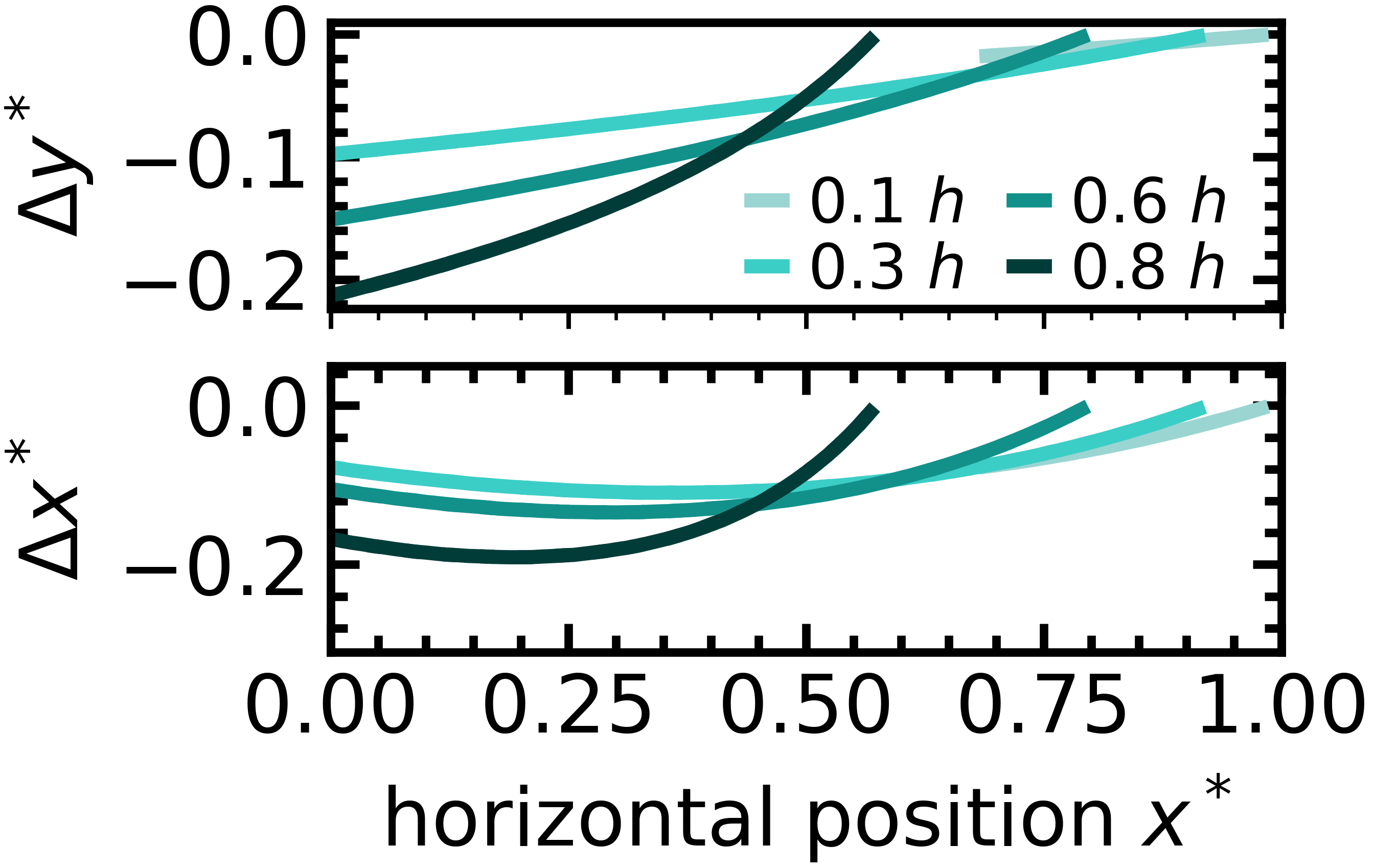}}
\captionof{figure}{Horizontal and vertical shifts in focus during horizontal measurement for a sessile drop at different heights. The positions marked in green in \figref{Fig: HorizontalRefraction} were evaluated. Scan paths closer to the substrate are shown in lighter shades of green.  \label{Fig: HorizontalRefractionCurves}}
\end{Figure} 
%
To validate the simulation results, the expected drop shape in the Raman image was calculated. The results for four droplets with different contact angles are shown in \figref{Fig: HorizontalDropletForm}. All the droplets show a V-shaped cut-out at the bottom. This is caused by the intersection of parts of the laser cone with the substrate and, as already described, leads to decrease in contrast compared to that predicted in the simulations but does not prevent the measurement. Owing to the small difference in curvature between the upper and lower laser cone rays, an extensive study of the composition at the \threepahsecontactline is possible for all contact angles. However, the positions of full focus close to the substrate deep within the droplet can only be reached for contact angles \ang{>=51} (see \figref{Fig: HorizontalRefraction}).
\begin{Figure}
\centerline{\includegraphics[width=180pt]{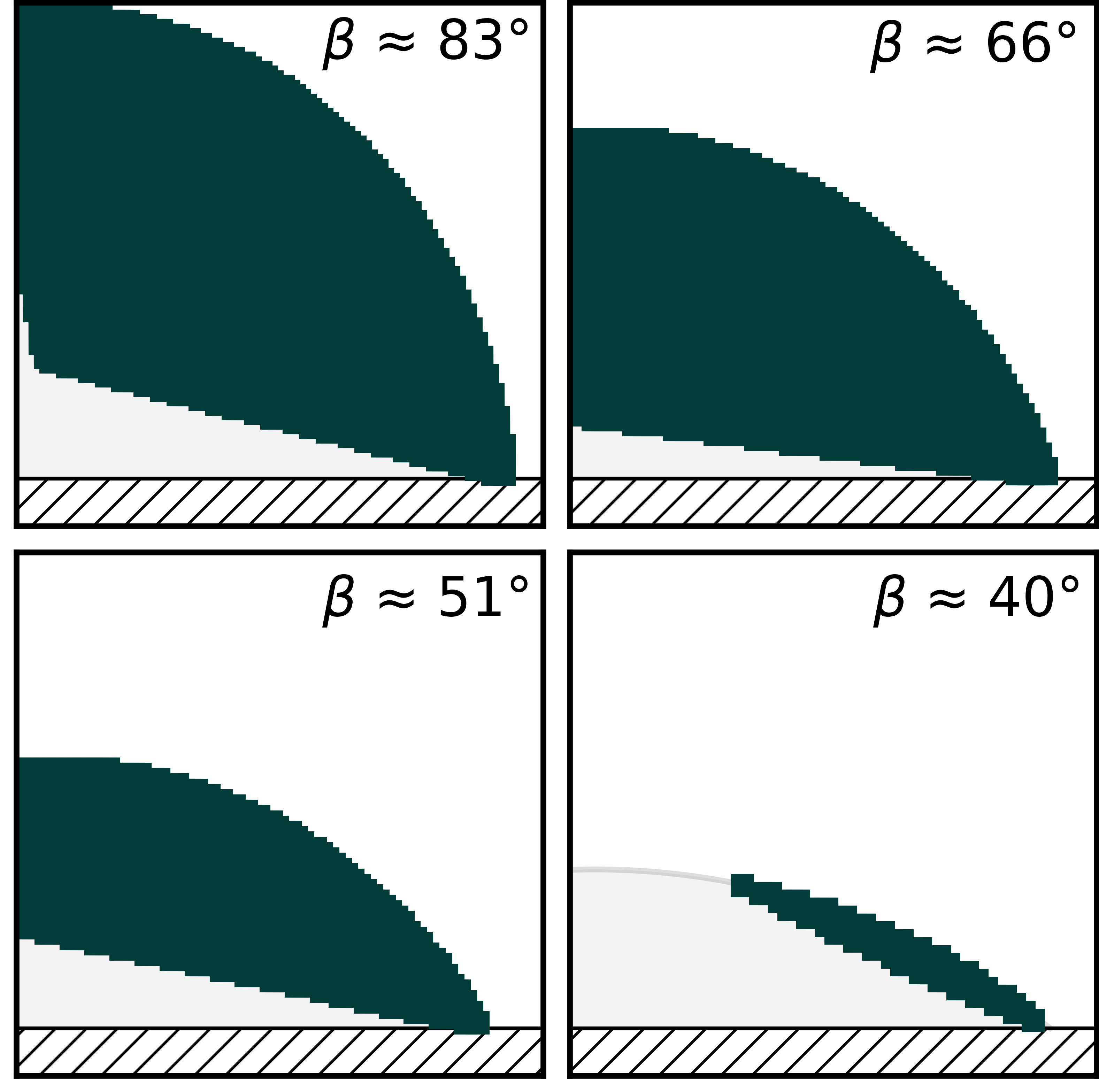}}
\captionof{figure}{Calculated drop shapes in the Raman image during a horizontal measurement. Focal points whose laser cones do not intersect the substrate are marked in green. The expected drop shape was evaluated for drops of different contact angles as shown in the top right corner. The drop shape predicted for $\contactangle=\ang{51}$ is similar to the drop shape measured in \figref{Fig: HorizontalGlyDroplet}.  \label{Fig: HorizontalDropletForm}}
\end{Figure}
%
Using horizontal Raman measurements, the concentration distribution near the \threepahsecontactline in an evaporating \qty{4.2}{\micro\liter} \qty{10}{mol\percent} glycerol/water drop was studied. In this fluid system, compositional changes are evident from variations of the OH and CH bands (see \figref{Fig: RamanSpectra}). The concentrations were calculated using a procedure analogous to that described in \cite{erb_visualization_2025}. Regions with high water concentrations are shown in yellow; regions with low water concentrations are shown in blue. Six measurements performed during the ongoing evaporation process are shown. The evaporation time is indicated on the left side of each concentration map. The drop contour can be observed via the contrast between yellow and blue at approximately 1/3 of the scan width. It shifts to the right as the evaporation time increases. The contact angle was approximately \ang{70}. Since a full-spectral evaluation method was used, the band of reduced intensity is not visible here. The scale bar shown at the bottom right applies to all measurements. Since water has a relatively high vapor pressure, it evaporates preferentially at the \threepahsecontactline. Thus, the water concentration decreases locally \cite{erb_visualization_2025, diddens_competing_2021}. As shown in \figref{Fig: HorizontalGlyWaterConcentrationMaps} as an expanding blue area, during evaporation, a region of low water concentration forms at the \threepahsecontactline. As expected, the water concentration at the \threepahsecontactline decreased from \qty{81.3}{mol\percent} to \qty{71.9}{mol\percent}. 
\begin{Figure}
\centerline{\includegraphics[width=200pt]{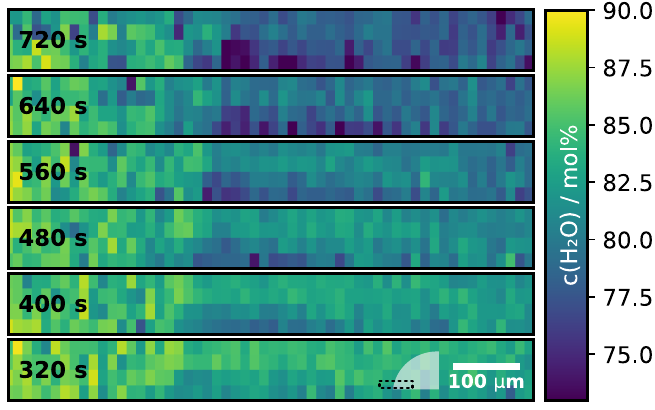}}
\captionof{figure}{Water concentration maps of the droplet region near the \threepahsecontactline. The mapped area relative to the droplet is shown as a black dashed rectangle in the lower right corner. Concentration maps of a \qty{4.2}{\micro\liter} \qty{10}{mol\percent} glycerol/water droplet evaporating at \qty{23}{\degree}C and a humidity of \qty{30}{\percent} measured after different evaporation times (as indicated on the left in each scan) are shown. Concentration maps were recorded using a 10x/0.25 objective, with a \qty{0.04}{\second} integration time, a laser power of \qty{17.5}{\milli\watt} and a resolution of \qtyproduct{22.5x14}{\micro\meter}. The water concentration was calculated by a procedure analogous to that described in \cite{erb_visualization_2025}. The calibration curve is shown in Fig. S2 on the SI. \label{Fig: HorizontalGlyWaterConcentrationMaps}}
\end{Figure}
\subsection{Discussion}
%
The shift in focus is directly proportional to the difference in curvature at the entry point of the outer laser cone rays. The greater the difference is, the larger the shift in focus. Thus, the shift in focus depends only on the drop shape and aperture angle of the laser cone but not on the drop size. Since the laser cone aperture angle increases with increasing numerical aperture, the refraction effects in the drop also strengthen (see Fig. S1 on the SI). The same applies to droplets with higher refractive indices. To reduce the shift in focus, Everall \etall suggested decreasing the optical density gradient by employing an immersion objective \cite{everall_confocal_2000, everall_modeling_2000}. However, this is not possible for a drop evaporating in a gaseous atmosphere.
\newline
In the simulations, only refraction effects were considered. However, in addition to refraction, diffraction might also occur inside droplets during Raman measurements. In this study, the confocal point of focus was calculated using geometrical optics as the point at which the outer laser cone rays intersect \cite{everall_confocal_2000, everall_modeling_2000}.  In confocal microscopy, the beam can be described by Gaussian beam optics \cite{su_simultaneous_2021}. As the focal point has a finite width in Gaussian beam optics, the beam waist deviates from the geometric laser cone, particularly near the focal point. Thus, when the focal point is only slightly beneath the drop surface, there can be slight deviations between the Gaussian focal point and the focal point calculated by using geometrical optics. Another possible source of error is the inhomogeneity of the refractive index. Since various studies have reported concentration gradients within droplets \cite{bell_concentration_2022, erb_visualization_2025, rostami_coalescence_2025}, a gradient in the refractive index within a droplet can also be expected. Nevertheless, the concentration gradients are typically small, so the assumption of a homogeneous index is justified.
\section{Conclusion}
In this study, we investigated the refraction of a Raman laser inside sessile droplets with contact angles \ang{<90} during Raman mapping. We calculated the shift in focus at different positions within the droplet for measurements with a vertically incident Raman laser as well as for measurements with a horizontally incident Raman laser and predicted the droplet shape in the Raman image for different contact angles. While for vertical measurements, the drop shape in the Raman image depends on the shift in the focal point, for horizontal Raman measurements, the overlap of the laser cone with the substrate leads to lower intensity bands at the lower end of the droplet.
\newline
The horizontal Raman measurements were enabled by strategically placing a \ang{45} mirror in the optical path of the laser. With this configuration, the concentration distribution near the \threepahsecontactline in an evaporating \qty{10}{mol\percent} glycerol/water droplet was resolved with an unprecedented resolution of \qtyproduct{14x22.5}{\micro\meter}. 
\newline
The findings of this study establish a framework for future Raman microscopic studies of sessile droplets. They thus contribute to our understanding of droplet dynamics, which has numerous practical applications in fields such as cooling, printing, and coating technology.

\subsection*{Author contributions}
Alexander Erb: Conceptualization, Methodology, Software, Validation, Formal analysis, Investigation, Data Curation, Writing - Original Draft, Writing - Review \& Editing, Visualization, Project administration. 
Johanna Steinmann: Investigation, Validation. 
Youngeun Lee: Investigation, Validation. 
Robert W. Stark: Conceptualization, Supervision, Funding acquisition, Writing - Review \& Editing.
\subsection*{Acknowledgments}
This study was funded by the Deutsche Forschungsgemeinschaft (DFG, German Research Foundation) – Project ID 265191195 – SFB 1194, “Interaction between Transport and Wetting Processes”, Project A07.
\subsection*{Financial disclosure}
None reported.
\subsection*{Conflict of interest}
The authors declare no potential conflict of interest.
%
\printbibliography
\end{multicols}

\end{document}


\maketitle

\section{Vertical Droplet scans with varying NA and contact angles}
To investigate the influence of the contact angle and the numerical aperture NA on the droplet shape distortion, \qty{3}{\micro\liter} glycerol droplets on different substrates were analyzed with two different objectives using a vertically incident Raman laser. The results are shown in \figref{Fig: DropletCrossSections}. To allow a comparison with the actual drop shape, the drops were additionally photographed with a side camera. The images are shown in the column on the right. The left column shows the CH integral of a measurement with a 10x/0.25 objective and the middle column shows the CH integral of a measurement with a 20x/0.40 objective. The top row shows droplets on Polydimethylsiloxane (PDMS) and the bottom row droplets on untreated glass. Regions of high integral are shown in yellow, regions of low integral are shown in blue. 
\begin{Figure}
\centering
\includegraphics[width=400pt]{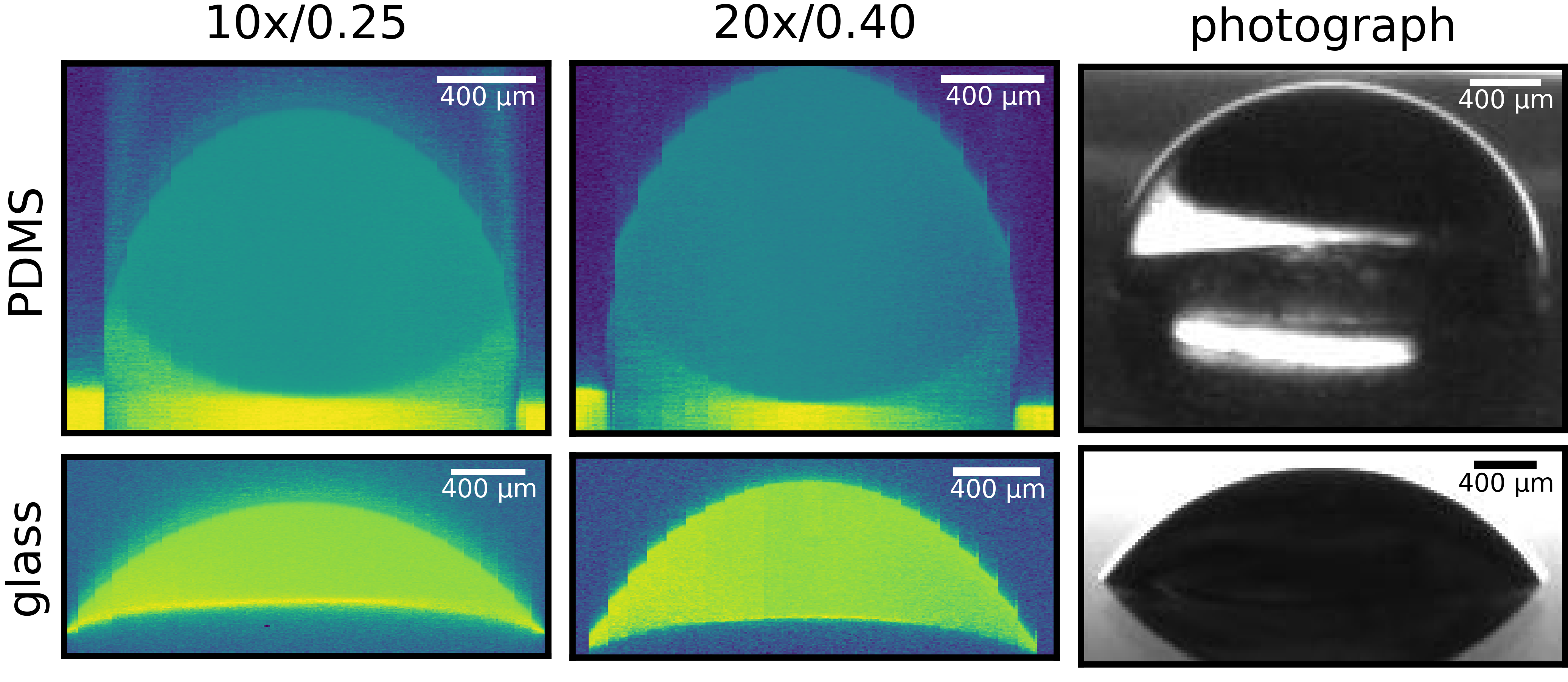}
\captionof{figure}{Caption \label{Fig: DropletCrossSections}}
\end{Figure}
\noindent
As PDMS is an organic material, it appears yellow in the CH integral images. The PDMS contact angle was found to be \qty{92(3)}{\degree}. As discussed in the manuscript, for droplets with high contact angles, the area around the \threepahsecontactline cannot be resolved due to refraction. The drop appears chopped off. With a larger aperture angle (larger NA), the shape distortion is more pronounced because the differences in slope of the two outer laser cone beams are greater. In addition, the droplet appears higher due to the stronger refraction effects in the center of the droplet.
\newline\newline
Also, on the glass substrate, the drop recorded with the large aperture angle appears higher. A contact angle of \qty{56(3)}{\degree} was measured. Since glass is an inorganic material, it is not visible in CH Integral image. As discussed in the manuscript, no convexity is visible in the center of the drop for small contact angles.
\section{Calibration for measurements on glycerol/water droplets}
To calculate water concentrations from the Raman spectra, the classical least squares method was used as described in \cite{erb_visualization_2025}. The calibration curve is shown in \figref{Fig: CalibrationCurve}.
\begin{Figure}
\centering
\includegraphics[width=150pt]{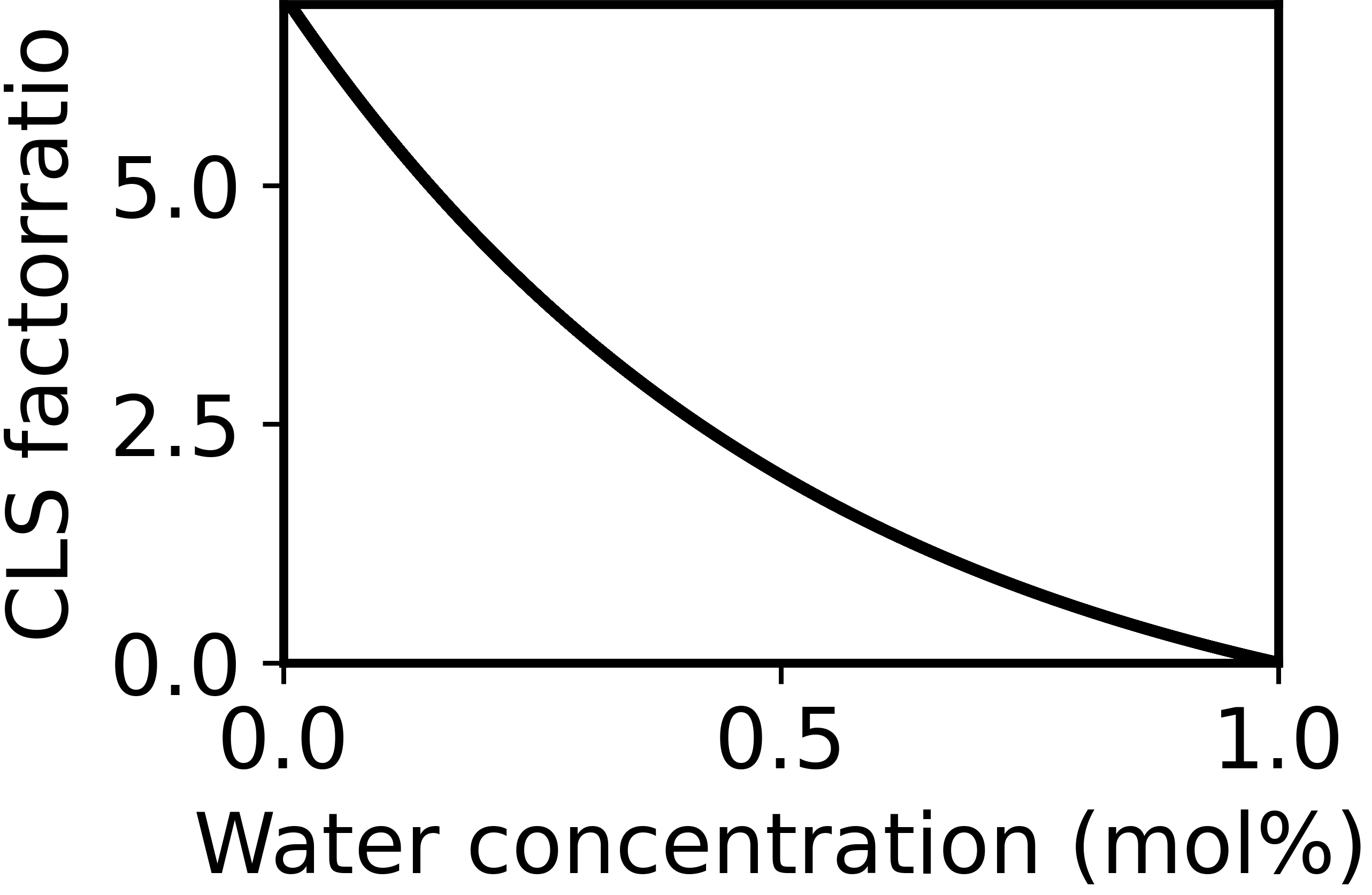}
\captionof{figure}{CLS Callibration curve as used to calculate water concentrations in the glycerol/water droplet. \label{Fig: CalibrationCurve}}
\end{Figure}

\printbibliography